# Securing SQLJ Source Codes from Business Logic Disclosure by Data Hiding Obfuscation

Praveen Sivadasan
PhD Scholar
School of Computer Sciences, MG University
Kottayam, Kerala, India
praveen_sivadas@yahoo.com

Dr.P.Sojan Lal
Research Guide
School of Computer Sciences, MG University
Kottayam, Kerala, India
sojanlal@gmail.com

*Abstract*— **Information security is protecting information from unauthorized access, use, disclosure, disruption, modification, perusal and destruction. CAIN model suggest maintaining the Confidentiality, Authenticity, Integrity and Non-repudiation (CAIN) of information. Oracle 8i, 9i and 11g Databases support SQLJ framework allowing embedding of SQL statements in Java Programs and providing programmer friendly means to access the Oracle database. As cloud computing technology is becoming popular, SQLJ is considered as a flexible and user friendly language for developing distributed applications in grid architectures. SQLJ source codes are translated to java byte codes and decompilation is generation of source codes from intermediate byte codes. The intermediate SQLJ application byte codes are open to decompilation, allowing a malicious reader to forcefully decompile it for understanding confidential business logic or data from the codes. To the best of our knowledge, strong and cost effective techniques exist for Oracle Database security, but still data security techniques are lacking for client side applications, giving possibility for revelation of confidential business data. Data obfuscation is hiding the data in codes and we suggest enhancing the data security in SQLJ source codes by data hiding, to mitigate disclosure of confidential business data, especially integers in distributed applications.**

**Keywords; Information Security, Data Obfuscation, SQLJ**

## I. INTRODUCTION

A stored procedure[10] is a program or routine that is invoked from a client program, and executes under the control of database manager[2,3]. The database programming tool SQLJ[1,11,12,13] which stands for "SQL-Java", allows SQL statements directly embedded within Java applications. DB2 and SQLJ support client side Java Stored Procedures[8], where the *Stored Procedure Specification* allows implementing database stored procedures and functions in Java. JDBC is a component that do not allow embedding of SQL statements[4], but allow SQL statements to be passed to the database and to be executed at the server side, for the performance of Client Server based Database applications. SQLJ and JDBC are defined in Java Source codes. The SQL statements are embedded in the Java program using the syntax #sql {<sql-statement>} and the SQLJ programs are with an extension of .sqlj.

SQLJ allows multiple connections to allow data access from several databases. It also allows to use Dynamic SQL.Also, PL/SQL stored procedures can be called from SQLJ, like #sql {CALL latestTransactionDate(:in m, :out lastDate)};. The syntax of SQLJ is compact than that of JDBC helping to write shorter programs with greater user productivity. It is Database neutral and works with Oracle, Sybase and DB2. The following shows samples of SQLJ and JDBC codes.

```
// SQLJ
int n;
#sql { INSERT INTO emp VALUES (:n)};

// JDBC
int n;
Statement stmt = conn.prepareStatement
         ("INSERT INTO emp VALUES (?)");
stmt.setInt(1,n);
stmt.execute ();
stmt.close();
```

During execution, the SQLJ applications are undergone through a Translation, Compilation and Customization process, in Figure 1.

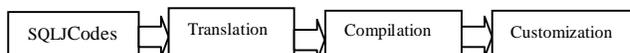

Figure1 SQLJ Execution steps

During Translation,in Figure2, the SQLJ program with an extension .sqlj is translated into a .java source file by the SQLJ Translator which automatically invokes the Java compiler to produce a .class file. The Java compiler generates byte codes for SQLJ and JDBC for the JVM. The Sematics-Checker or Checker performs the translation time syntax and semantics checking of the SQL statements. The translation, in Figure3, would result in output files with extension .java, followed by compilation create files with extensions. class and .ser. The .class and .ser files are archived into a .jar file.

The .java files are calls to the SQLJ runtime, which in turn calls the JDBC driver during application execution. The serialized resource files with extension .ser contain the application profiles which are information about all the SQL instructions. In customization, the profile created during the translator's code generation phase is customized for use with particular database and performance enhancements specific to Oracle databases.

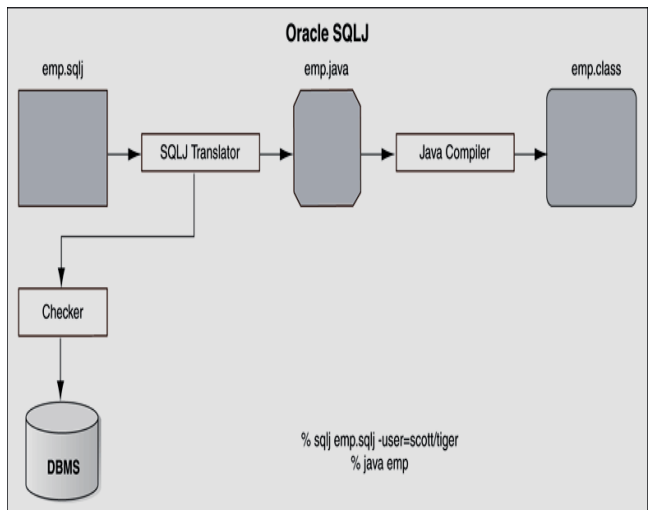

Figure 2 SQLJ Translator Framework

Oracle JDeveloper10g is a Java-based Integrated Development Environment(IDE) for SQLJ programming.

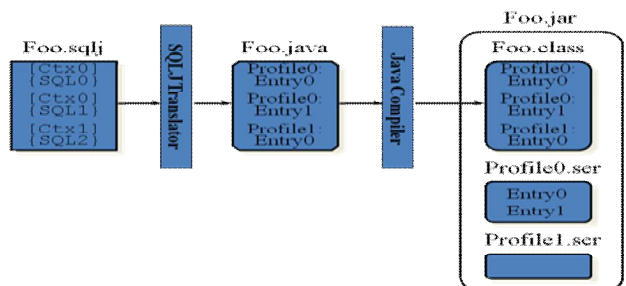

Figure 3 Output files after translation and compilation

## II. SOFTWARE OBFUSCATION

Software obfuscation or code obfuscation [5,6] transforms a program into an obfuscated program which is difficult to comprehend. The various popular obfuscation transformation techniques are Lexical Transformation, Data Transformation and Control Transformation. Layout transformation makes the structure of the transformed program difficult to comprehend, by replacing the mnemonics of original structure with unreadable formats. Data transformation obscures the crucial data and data structures of the original program, whereas control transformation obscures its flow of execution. Data Hiding is a data transformation technique that hides the data of the original program. The effectiveness of obfuscation is usually measured in terms of a) the *potency* that is the degree to which the reader is confused[9] b) the *resilience* that is the degree to which the obfuscation attacks are resisted and finally c) the *cost* which measures the amount of execution time/space penalty suffered by the program due to obfuscation.

The source code and byte code can be obfuscated and despite the fact that implementation of source code obfuscators is little hard, the obfuscator can apply transformations to codes at any stage of development. It also adds significant strength to byte codes and thereby making the business data also hidden in the decompiled version.

Encrypting of code makes it mandatory to decrypt it prior to execution, which leads to interception and stealing of codes by special hardwares. But obfuscation helps the code to execute without transforming to its original format making it better than cryptographic approaches to add more strength to software codes. Deobfuscators are tools that generate original codes from obfuscated versions. Reverse engineering effort is the effort for the malicious reader in understanding obfuscated codes.

## III. DISCLOSURE OF INTEGER DATA IN SQLJ CODES VIOLATING CONFIDENTIALITY

Various customized or internet banking applications are developed in Java allowing EBusiness operations to be carried out efficiently between banks and other enterprises. SQLJ is a popular tool that allows SQL and PL/SQL stored procedures to be embedded in Java and provides high flexibility for Java based database applications development.

The PL/SQL stored procedures and the database contents at the server side are well protected by encryption and strong security techniques[14]. But, the SQLJ procedures at the client side could surely expose business operation details to a malicious reader.

We realized that the SQLJ procedures can be strengthened and protected from revelation of business logic by obfuscation than encryption, allowing maximum Java Stored Procedures definitions and its utilization. We propose strengthening of SQLJ embedded stored procedures by data hiding, especially by hiding the crucial and confidential integer business data in procedures.

In the following paragraphs, we present some sample SQLJ codes that expose integer business data.

**Case1:**// SOURCE FILE NAME: TbSel.sqlj, Copyright IBM Corp. 2007
   /* The following if statement sets the new employee's benefits based on their years of experience.
   */ if(years > 14)

benefits = "Advanced Health Coverage and
Pension Plan";
            else if(years > 9)
              benefits = "Advanced Health Coverage";
            else if(years > 4)
              benefits = "Basic Health Coverage";
            else
         benefits = "No Benefits";

**Case2:** // SOURCE FILE NAME: SpClient.sqlj
        // set the input parameters of the stored procedure
    inParamLowSal = 15000;
    inParamMedSal = 20000;
    inParamHighSal = 25000;
    System.out.println("Call stored procedure named " + procName);
    #sql {CALL IN_PARAMS(:in inParamLowSal, :in inParamMedSal,
                :in inParamHighSal, :in inDept)};

    System.out.println(procName + " completed successfully");

    // display total salary after calling IN_PARAMS
    #sql {SELECT SUM(salary) INTO :sumSalary
         FROM employee
         WHERE workdept = :inDept};

**Case3:** // SOURCE FILE NAME: TbCreate.sqlj
// create a table called emp_details with a 'CHECK' constraint
    System.out.println(
      "\n  CREATE TABLE emp_details(empno INTEGER NOT NULL PRIMARY KEY,\n" +
      "           name VARCHAR(10),\n" +
      "           firstname VARCHAR(20),\n" +
      "           salary INTEGER CONSTRAINT minsalary\n" +
      "             CHECK (salary >= 25000)\n" +
      "             NOT ENFORCED\n" +
      "             ENABLE QUERY OPTIMIZATION)\n");
    #sql {CREATE TABLE emp_details(empno INTEGER NOT NULL PRIMARY KEY,
               name VARCHAR(10),
               firstname VARCHAR(20),
               salary INTEGER CONSTRAINT minsalary
                 CHECK (salary >= 25000)
                 NOT ENFORCED
                 ENABLE QUERY OPTIMIZATION)};
" Invoke the statement:\n" +
      "    INSERT INTO staff(id, name, dept, salary)\n" +
      "    SELECT INTEGER(empno)+100, lastname, 77, salary\n" +
      "    FROM employee\n" +
      "    WHERE INTEGER(empno) >= 310");

    #sql {INSERT INTO staff(id, name, dept, salary)
         SELECT INTEGER(empno) + 100, lastname, 77, salary
         FROM employee
         WHERE INTEGER(empno) >= 310};

#sql {UPDATE staff
         SET salary = salary + 1000
         WHERE id >= 310 AND dept = 84};

**Case4:** // SOURCE FILE NAME: TbCursor.sqlj
    // display the final content of the 'staff' table
    staffTbContentDisplay();

    System.out.println();
    System.out.println(
      "  Invoke the statements:\n" +
      "    DECLARE posCur CURSOR FOR\n" +
      "      SELECT name, dept FROM staff WHERE id >= 310\n" +
      "\n" +
      "    FETCH :posCur INTO :name, :dept\n" +
      "    while (successful fetch)\n" +
      "    {\n" +
      "      if (dept != 84)\n" +
      "      {\n" +
      "        UPDATE staff\n" +
      "        SET comm = 0.01 * (SELECT AVG(salary)\n"+
      "                FROM staff\n" +
      "                WHERE id >= 310)\n" +
      "        WHERE CURRENT OF :posCur\n" +
      "      }\n" +
      "      FETCH :posCur INTO :name, :dept\n" +
      "    }");
    // declare a cursor
    #sql posCur = {SELECT name, dept FROM staff WHERE id >= 310};

The above discussed codes contain integers that could expose the business logic to a malicious reader and we suggest hiding the data, to maintain the data confidentiality in the source code.

## IV. PROPOSAL TO HIDE INTEGERS IN SQLJ CODES

Our Data hiding transformation considered and discussed in [7], hides integer value 2 by an F(a,b) call. Any integer say 'm' is represented as (m/2)*F(c,d),if m is even and (m/2)*F(c,d)+1,if m is odd. Each successive iteration further tries to hide the parameter 'c' of the F() calls, making it nested like F(F(F(F()))), making the reader hard to find the hidden value. A sample java code is shown below to explain the data obfuscation process.

The file Query.java is as follows.

```java
class Query
{ public static String query;
public static String Obfus( )
{query="UPDATE Account_Holders
SET Acct_Balance=Acct_Balance+"+
Integer.toString(50)+
" WHERE  Acct_Balance>"+Integer.toString(5000);
            return(query);}}
```

The following is an integer hidden obfuscated code.

```java
class Query_mod123123123123 extends obfuscate
{
public static String query;
public static String Obfus( )
{query="UPDATE          Account_Holders          SET
Acct_Balance=Acct_Balance+"+Integer.toString((25*F(F(F(F
(1524%767,7)*187%191,5)*762%767,7)*187%191,5)))+
WHEREAcct_Balance>"+Integer.toString((2500*F(F((F((F(4
9135%24575,12)*378+F(33%21,2))%383,6)*24567+F(33%2
1,2))%24575,12)*3065%3071,9)));
return(query);}}
```

## CONCLUSION

The data hiding adds significant reverse engineering effort and potency to the codes, without costing much on its execution time. It is convinced that data hiding obfuscation maintains the confidentiality of application data, conforming to the CAIN model. We are in the initial implementation phase of SQLJ obfuscator and it has been named as "JSQLJ Obfuscator".